\documentclass[aps,pre,showpacs, twocolumn]{revtex4}

\usepackage{graphicx}
\usepackage{amssymb}
\usepackage{amsmath}
\usepackage{colordvi}
\usepackage{color}

\newcommand{\rev}[1]{\textcolor{black}{#1}}
\newcommand{\cH}{{\cal H}}
\newcommand{\cI}{{\cal I}}

\begin{document}

\title{Metastable soliton necklaces confined by the boundary of a flattop region}

\author{Dmitry A. Zezyulin}

\affiliation{School of Physics and Engineering, ITMO University, St. Petersburg 197101, Russia}
\affiliation{Bashkir State Pedagogical University n.a.  M. Akhmulla, Oktyabrskoj rev. str., 3a, 450077, Ufa, Russia}

\date{\today}

\begin{abstract}

We present quasistationary ring-shaped soliton necklaces in a two-component envelope propagating in a medium with competing cubic-quintic nonlinearity. Metastable propagation of soliton necklaces results from a balance of repulsion between adjacent out-of-phase solitons in one component  and confinement by the boundary of a flattop region in the other. Numerical simulations demonstrate metastable propagation over \rev{about a hundred} diffraction lengths, even with random noise added to the input envelopes. The maximum  number of solitons in metastable necklaces can be controlled either by changing the size of individual solitons or by adjusting the width of the flattop region hosting the necklace.

\end{abstract}

\maketitle
 
 A soliton necklace is an array of closely spaced solitons arranged circularly   in the transverse plane  \cite{Soljacic98}. In general,  the necklace beam does not propagate in a stationary manner, as the complex interplay of inter-soliton forces gives rise to a variety of   dynamical behaviors,  ranging  from the fusion of individual solitons   to the unbounded expansion of the necklace  radius. In addition, in a   self-focusing Kerr medium, the necklaces can be unstable against the collapse of individual solitons   \cite{Grow07}. The creation of stationary (or at least quasi-stationary) soliton necklaces remains a challenging issue that has received much attention. Various strategies have been proposed to promote steady evolution of necklace-shaped soliton clusters and enhance their robustness to perturbations. These include phase imprinting inducing a nonzero net angular momentum \cite{Soljacic01,Desyatnikov02},  the use of quadratic \cite{KartashovJOSAB02} and nonlocal nonlinearities \cite{Buccoliero07},    competing quadratic-cubic  \cite{KartashovPRL02} or cubic-quintic nonlinearities  \cite{Mihalache2003},  and confinement by external trapping potentials, which can also be combined with  a quintic \cite{Dong23} or saturable nonlinearity \cite{Lang25}. Metastable soliton necklaces have also been reported to exist in the presence of fractional diffraction \cite{Li20}. In experiments, enhanced stability of annular soliton clusters and multipole solitons has been confirmed   in nonlocal \cite{Rotschild06} and cubic-quintic media \cite{Reyna20}. 
 
 A particularly promising approach to   robust necklace-shaped soliton clusters involves vectorial models that describe the propagation of multiple incoherent beams. Effective attractive interactions between the components can slow the expansion of the necklace as compared to the scalar case \cite{DesyatnikovPRL01}. A solitary wave forming in one component can create an effective waveguide that stabilizes multipole solitons forming in the other component \cite{DesyatnikovOL01}. Composite solitons of this type have been experimentally created in a nonlinear saturable  medium \cite{DesyatnikovOL01,Desyatnikov02,Desyatnikov_JOSA02}.
 
 The main goal of this Letter is to elucidate a distinct mechanism that enables quasistationary propagation of multisoliton necklaces in a medium with two incoherently coupled envelopes and competing cubic-quintic nonlinearity. We show that the expansion of an out-of-phase soliton necklace forming in the first component can be compensated by the confining action of the boundary of the broad flattop region in the second component. As a result of this interplay, a metastable necklace can emerge.   
 
 \rev{The confinement by the boundary of the flattop region can be viewed as a form of nonlinear total internal reflection that prevents the individual solitons forming the necklace from escaping the flattop intensity region \cite{Zezyulin26}.} In the liquid light analogy \cite{liquid}, this effect can be understood as surface tension associated with the interface of a light droplet \cite{surface,theory}. The stabilization mechanism is thus  physically transparent and explicitly exploits the unique combination of properties inherent to bimodal envelopes propagating in media with competing nonlinearities.
 
 We present our results using the  normalized model for a two-component envelope $(\psi_1, \psi_2)$   propagating along the $z$-direction:
 \begin{equation}
 \label{eq:main}
 i\frac{\partial \psi_{1,2}}{\partial z}  +\frac{1}{2} \left(\frac{\partial^2 \psi_{1,2} }{\partial x^2} + \frac{\partial^2  \psi_{1,2}}{\partial y^2}\right)    -  \frac{\partial \cH(\cI_1, \cI_2)}{\partial \cI_{1,2} } \psi_{1,2} = 0,
 \end{equation}
 where  $(x,y)$ are the transverse coordinates,   and the function  $\cH(\cI_1, \cI_2)$, where   $\cI_{1,2} = |\psi_{1,2}|^2$,  describes  nonlinear  self-action and incoherent coupling between the components.  We assume 
 \begin{equation}
 \label{eq:Hamilt}
 \cH = -{(\cI_1^2 +  \cI_2^2)}/{2}  + {(\cI_1^3  + \cI_2^3)}/{3}   - \beta \cI_1 \cI_2,
 \end{equation}
 which corresponds to competing cubic-quintic nonlinearity in each component and effective attraction between the components for  $\beta>0$. The model defined by  (\ref{eq:main})--(\ref{eq:Hamilt}) was introduced in \cite{Maimistov99} to describe incoherent attraction between two-dimensional solitons in a bimodal system. This model and its modifications have been used to study two-component vortex solitons \cite{Mihalache2002,Mihalache_PRE2002,Mihalache2004,Desyatnikov2005}. 
 Below, we assume $\beta=2$, which corresponds to  two pulses propagating with distinct carrier frequencies. In optical experiments, competing cubic-quintic media were implemented in the polydiacetylene \emph{para}-toluene sulfonate \cite{Lawrence98}, liquid carbon disulfide \cite{Reyna20}, silver colloids \cite{colloids}, and in a coherently prepared multilevel atomic medium  \cite{atoms}. \rev{In physical units, propagation over $z = 1$ corresponds to the distance $Z = \lambda |n_3| / (2\pi n_2^2)$, where $\lambda$ is the wavelength, and $n_2>0$ and $n_3<0$ are the nonlinear refractive-index terms corresponding to cubic and quintic nonlinearity, respectively. Adopting, for reference,  values from \cite{Lawrence98}, we estimate $Z \approx  0.042$ mm. Thus, one  diffraction length $L_D = 0.85$~mm corresponds to $z \approx 20$.}
 
\begin{figure}
 	\begin{center}		
 		\includegraphics[width=0.999\columnwidth]{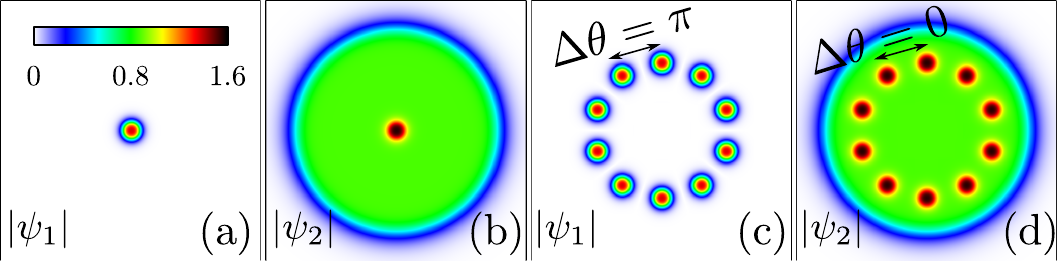}	
 		\caption{The amplitude distributions $|\psi_{1,2}(x,y, z=0)|$ for the    seed solution (a,b)  with propagation constants $(b_1, b_2) \approx (3.0525, 0.1795)$ and for  the   soliton necklace (c,d) with $N=10$,  $M=5$, and $R=13$, generated from this seed.  The next-neighbor solitons in the necklace are out-of-phase in the first component (i.e., $\Delta \theta=\pi$), and in-phase in the second (i.e., $\Delta \theta = 0$).  All plots  are shown within the transverse spatial window $x, y \in [-25,25]$. The   colorbar applies to all panels.\label{fig:seed} }
 	\end{center}
 \end{figure}

 We construct a multisoliton necklace from a superposition of several copies of a seed solution.  To find the seed solution, we use the substitution  $\psi_{1,2} = e^{i b_{1,2} z} u_{1,2}(x,y)$, where $b_{1,2}$ are real propagation constants, and $u_{1,2}(x,y)$ are real functions. For certain combinations of $b_1$ and $b_2$, there exist solutions $u_{1,2}(x,y)$ with essentially different intensity distributions \cite{Zezyulin26}: the first component is a solitary wave on a zero background, while the second component takes the form of a similar solitary wave sitting on a broad flattop plateau. \rev{In these solutions, $b_1$ controls the size of the solitary wave, while $b_2$ controls the width of the flattop region and therefore  must remain close to the cutoff value $3/16 = 0.1875$ \cite{surface,theory,Zezyulin26}.} An example of a seed solution   and a soliton necklace created from this seed are shown in Fig.~\ref{fig:seed}. 
 
 To create the soliton necklace, we represent the second component of the  seed solution in the form  $u_2(x,y) = u_2^{FT}(x,y) + v_2(x,y)$, where $u_2^{FT}$ describes the broad flattop profile, and $v_2$ is the narrower solitary wave situated in the center of this plateau. The necklace of $N$ solitons is then generated by placing $N$ copies of $u_1$ and $v_2$ along a circle:
 \begin{eqnarray}
 \label{eq:psi1}
 \psi_1(x,y, z=0) = \sum_{k=1}^N u_{1}(x-x_k, y-y_k) e^{2\pi i M k/N},\\
 \label{eq:psi2}
 \psi_2(x,y, z=0)  = u_2^{FT}(x,y) +  \sum_{k=1}^N v_2(x-x_k, y-y_k),
 \end{eqnarray}
 where $(x_k, y_k)  = R(\cos(2\pi k/N), \sin(2\pi k/N))$ is the position of the center of the $k$th solitary wave in the necklace, and $R$ is the necklace radius. The integer $M$ is a free parameter that controls the phase difference  between  adjacent solitons in the first component. Without loss of generality, it is sufficient to consider $0\leq M \leq N/2$.  The phase difference between  adjacent solitons is zero for    $M=0$ and $\pi$ for $M=N/2$. For $0<M \leq N/4$ and $N/4<M\leq N/2$ the interactions between adjacent solitons in the first component are attractive and repulsive, respectively. There is no phase difference between   individual solitons in the second component.
 
 The existence of robust soliton necklaces can be anticipated from the following  reasoning. For  repulsive interactions between adjacent out-of-phase solitons in the first component, the necklace formed in this component tends to expand.   This expansion is transferred to the second component via the  attractive coupling.  However, the surface tension of the flattop light droplet in the second component is expected to prevent  indefinite growth of the necklace radius. Thus, a (quasi)stationary pattern can emerge from the balance between the repulsive forces in the first component and the confining effect in the second. 
 
\begin{figure}
 	\begin{center}		
 		\includegraphics[width=0.999\columnwidth]{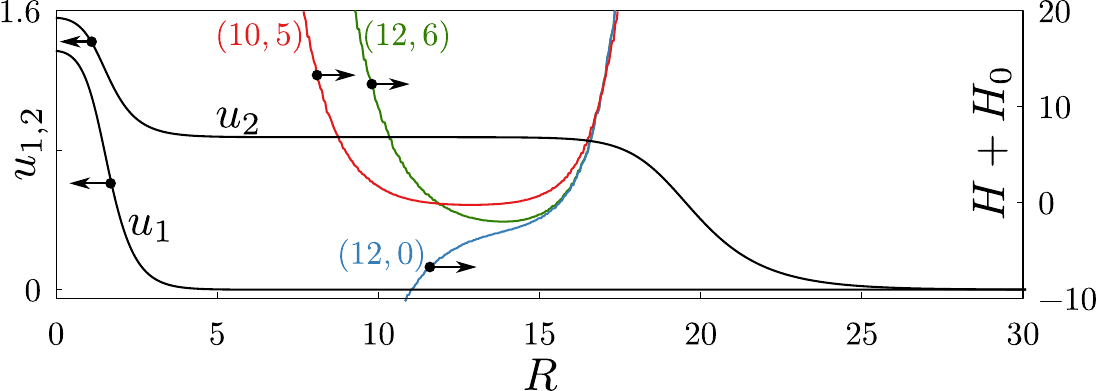}	
 		\caption{Radial profiles $u_{1,2}(R)$ of the seed solution from Fig.~\ref{fig:seed}  (left axis) and several dependencies $H(R)$  for different combinations $(N,M)$ (right axis). For better comparison,  each dependence $H(R)$ is shown up to an irrelevant constant shift $H_0$, which differs for each   pair   $(N,M)$. \label{fig:H} }
 	\end{center}
 \end{figure}

 In support of  this argument, we examine the Hamiltonian
 \begin{equation}
 H = \iint_{\mathbb{R}^2}\left[  \frac{1}{2}(|\nabla \psi_1|^2 + |\nabla \psi_2|^2)+  \cH(|\psi_1|^2, |\psi_2|^2)\right]dxdy.
 \end{equation}
 The value of  $H$ is conserved along the propagation distance $z$.  Stable solutions correspond to minima  of  $H$ achieved under the condition of fixed beam power $P= \iint_{\mathbb{R}^2} (|\psi_1|^2 + |\psi_2|^2)dxdy$. We fix all parameters except for the necklace radius $R$ and examine the function $H(R)$. Three   plots  of $H(R)$ are  shown  in Fig.~\ref{fig:H}. In agreement with the above reasoning, for $M=N/2$ the dependencies $H(R)$ have their minima for $R$ lying close to the boundary of the flattop region. For comparison,  we also plot an analogous  dependence with $M=0$, i.e., when  all solitons are in-phase. No minimum of $H(R)$ is achieved in this case.
 
 It is expected that the highest chance of obtaining stable soliton necklaces corresponds to radii that minimize the dependencies $H(R)$. In order to verify this, we use the distributions (\ref{eq:psi1})--(\ref{eq:psi2}) as initial conditions and simulate their propagation in (\ref{eq:main}). We compute the radius corresponding to the position of the maximal intensity in the second component: $R_{m}(z) = (x_{m}^2(z) + y_{m}^2(z))^{1/2}$, where $ (x_{m}(z), y_{m}(z)) = \mathrm{argmax}_{(x,y)}\, |\psi_2(x,y,z)|$. As long as the solution preserves the circular necklace shape, the dependence $R_m(z)$ is nearly constant and  coincides with the analogous characteristic  computed from the first component. 
 
\begin{figure}[t]
 	\begin{center}		
 		\includegraphics[width=0.999\columnwidth]{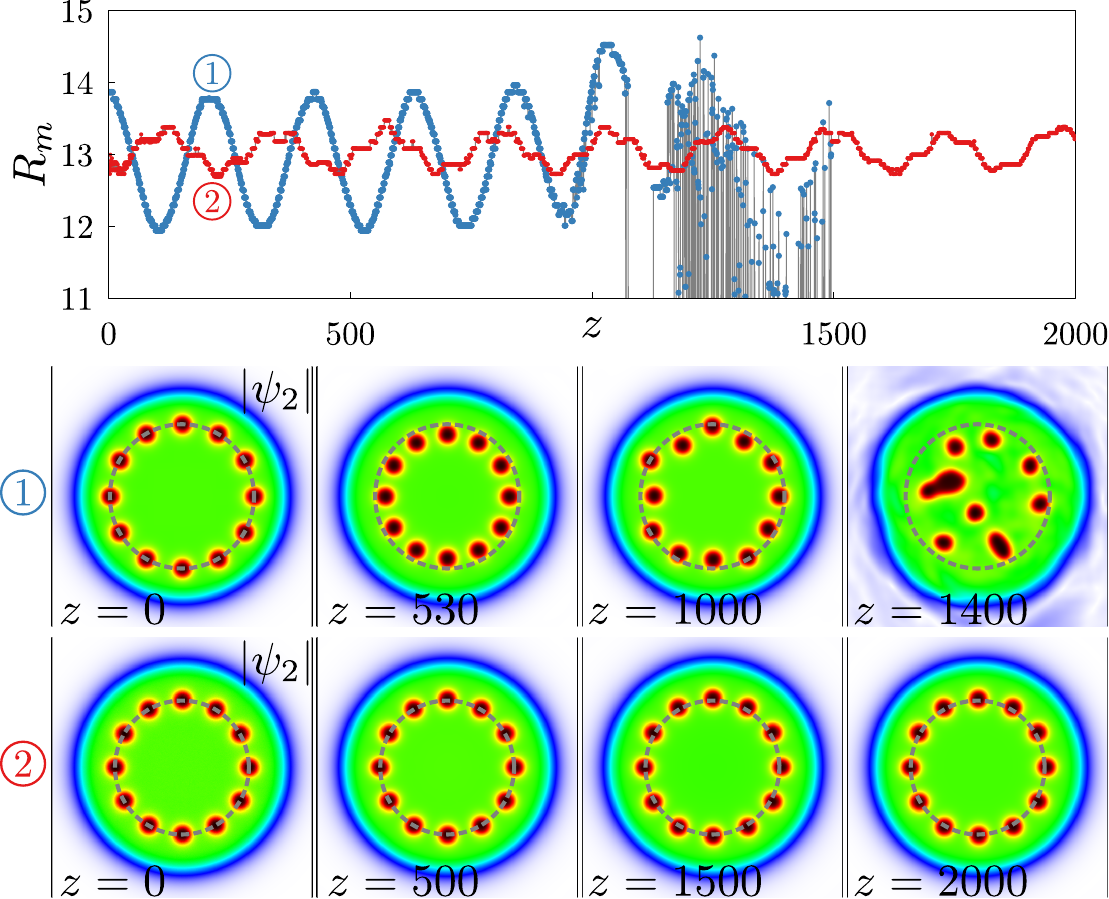}	
 		\caption{The upper panel shows the dependence $R_m(z)$, for the soliton necklace initialized with radius $R$  minimizing the dependence $H(R)$ (curve~1) and with  the mean oscillation radius (curve~2). The latter case includes  additional random noise in the input solution. The calculation corresponding to curve~1 was terminated at $z=1500$. Second and third rows show the amplitude distributions of $|\psi_2|$ at different  $z$ for curve~1 (the second row) and curve~2 (the third row). Amplitude distributions  $|\psi_1|$ are  similar,  except that the flattop plateau is absent in   $\psi_1$. All plots in the second and third rows are shown within the transverse spatial window $x, y \in [-25,25]$. Gray dashed circles show the  radii of the  input necklaces: $R(z=0)=13.90$ (curve~1) and $R(z=0)=12.91$ (curve~2). Colorbar as  in Fig.~\ref{fig:seed}. 
 			\label{fig:N12}}
 	\end{center}
 \end{figure}

 Systematic numerical simulations for different $N$ and $M$ indicate that the  input radii $R$ that minimize  $H(R)$ typically slightly  overestimate  the optimal necklace radii for the most robust propagation. \rev{A possible reason for this may be that when the individual solitons are brought closer to the boundary of the flattop region, their shapes could adjust accordingly --- an effect that would not be captured by the simplified model based on the $H(R)$ dependence, which assumes that the shape of the individual solitons is independent of $R$.} As a result, a necklace initialized with a minimizing radius initially tends to shrink. However, once the necklace radius becomes sufficiently small, the repulsion between adjacent out-of-phase solitons strengthens, causing the necklace to expand back toward its initial radius. This interplay gives rise to an oscillatory regime that typically persists only over a few periods, after which the necklace deforms from its circular shape and is destroyed. An example of propagation of this type corresponds to curve~1 in Fig.~\ref{fig:N12}. \rev{Next, we use the first few stable oscillation periods to compute the mean oscillation radius and create an input necklace initialized with this radius.}  When we simulate the propagation of this necklace, we observe that the oscillations of its radius are suppressed and its stability is dramatically enhanced. Moreover, in this case the robust propagation persists even when the initial distributions are additionally perturbed by random complex-valued noise whose maximum amplitude is a few percent of the solution amplitude. The soliton necklace can propagate virtually undistorted for thousands of \rev{units of $z$}, as shown by curve~2 in Fig.~\ref{fig:N12}.    \rev{Propagation over $z = 2 \times 10^3$  corresponds to approximately one hundred diffraction lengths. Comparing this result with earlier studies, where soliton clusters were deemed stable and experimentally feasible if they propagated over several tens of diffraction lengths \cite{Soljacic98,Soljacic01,Desyatnikov02,Mihalache2003,Reyna20,Desyatnikov_JOSA02}, the obtained  solution can also be considered metastable.} 
 
 \rev{The   perturbation of the input necklace was implemented as
 	\begin{equation}
 	\psi_{1,2}(x,y,0) \to \psi_{1,2}(x,y,0)\left[1 + r\left(p_{1,2}(x,y) + i q_{1,2}(x,y)\right)\right],
 	\end{equation}
 	where the  coefficient $r$ characterizes the perturbation strength, and $p_{1,2}$ and $q_{1,2}$ are obtained from random noise. In the results presented hereafter, we set $r = 0.02$ and used the MatLab function \texttt{rand} to generate pseudorandom numbers.
 }
 
\begin{figure}
 	\begin{center}		
 		\includegraphics[width=0.999\columnwidth]{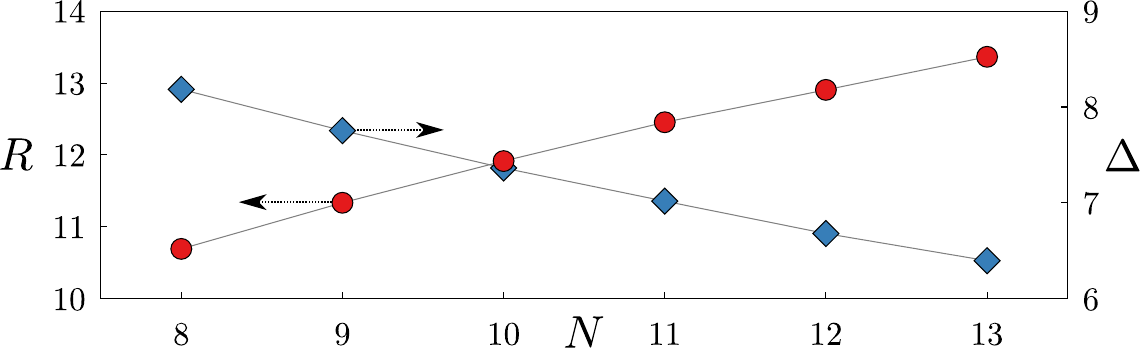}	
 		\caption{The radius $R$ of the most robustly propagating necklace (left axis) and the distance $\Delta$ between the next-neighbor solitons (right axis), both  plotted versus the soliton number $N$.  For each $N$, the  necklace was generated with $M= \mathrm{floor}( N/2)$. \label{fig:R} }
 	\end{center}
 \end{figure}

 Our calculations indicate that, for the given seed solution, the best stabilization of soliton necklaces can be achieved if $N$ is not too large. For the seed soliton solution used above, the maximal possible $N$ is equal to $12$. For necklaces with $N=13, 14, \ldots$ quasistationary propagation can only be sustained  for hundreds of units of $z$.   At the same time, there are metastable necklaces with $N=11, 10, 9, \ldots$. The most robust solutions are, as expected,  obtained for even $N$ and $M=N/2$, when  adjacent solitons in the first component are exactly out of phase. When the condition $M=N/2$ is not satisfied, the solutions become less robust, which can be explained by weaker repulsive interactions between individual solitons. In addition, any  input necklace beam with  $0<M<N/2$  carries a nonzero angular momentum \cite{Soljacic01}, which leads to the rotation of the necklace as a whole and  thus accelerates   the instability. Nevertheless, even when $M\ne N/2$, metastable propagation was observed for the necklaces   with $(N,M) = (12, 5), (11, 5), (9,4)$, etc.  The necklace radii corresponding to the most robust propagation increase with  $N$, as plotted in Fig.~\ref{fig:R}. The distance between the adjacent solitons, $\Delta := R\sin(2\pi/N)$ decreases as  $N$  and $R$ increase.  This is because larger-radius necklaces experience stronger confinement from the boundary, which forces the solitons closer together. The reduced spacing in turn strengthens their mutual repulsion, which is needed to counterbalance the increasingly tight confinement.
 
\begin{figure}[t]
 	\begin{center}		
 		\includegraphics[width=0.999\columnwidth]{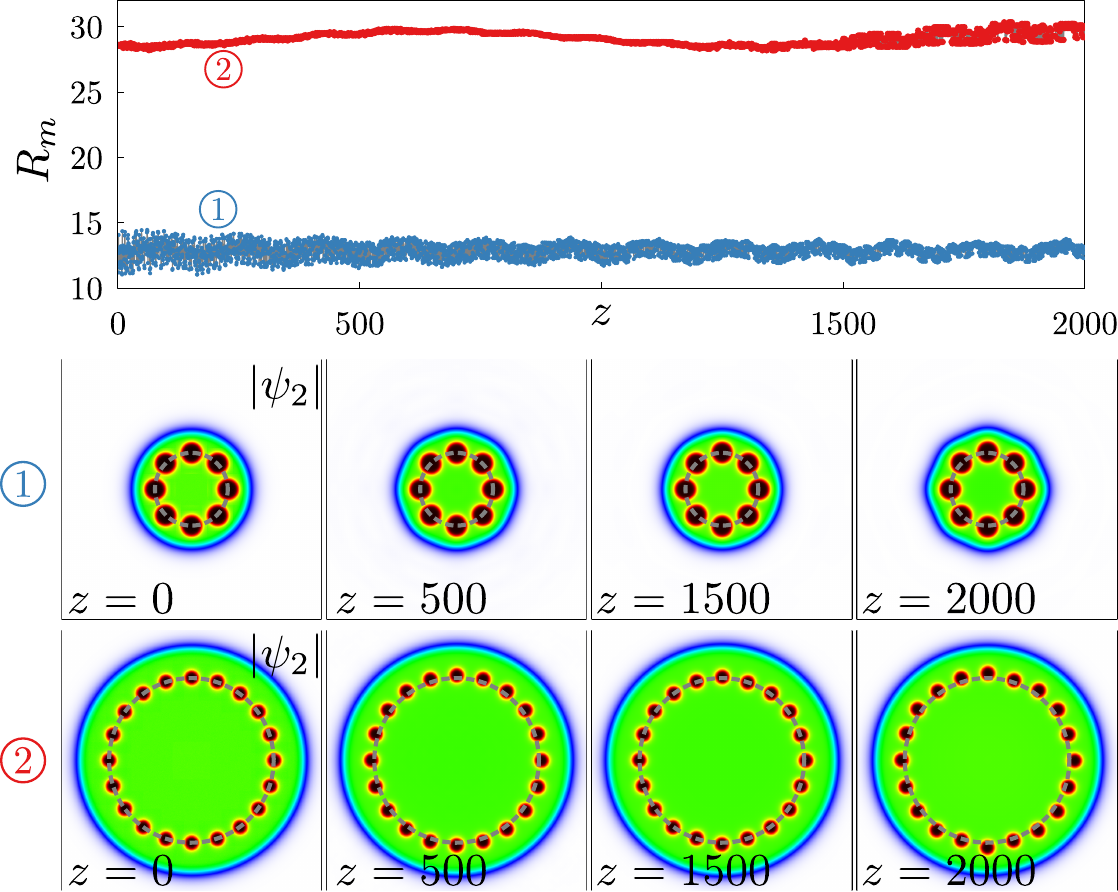}	
 		\caption{The upper panel shows the dependence  $R_m(z)$ for the soliton necklace with $(N,M)=(8,4)$ generated from the seed solution with the propagation constants $(b_1, b_2)\approx (2.1225, 0.1795)$ (curve~1) and for the necklace  with $(N,M) = (20,10)$   generated from the seed solution with   $(b_1, b_2)\approx (3.1825, 0.1835)$ (curve~2).  Both simulations include  additional random noise in the input solutions.  Second and third rows show the amplitude distributions of $|\psi_2|$ at different   propagation distances $z$ for  curve~1 (the second row) and curve~2 (the third row).   All plots in the second and third rows are shown within the transverse spatial window $x, y \in [-45,45]$. Gray dashed circles show the  radii of the  input necklaces:  $R(z=0)=12.64$ (curve~1) and $R(z=0)=28.56$ (curve~2).  Colorbar as  in Fig.~\ref{fig:seed}.   \label{fig:long} }
 	\end{center}
 \end{figure}

 The number of solitons in metastable necklaces can be controlled either by the width of the individual solitons in the seed solution  or by the radius of the flattop plateau. These characteristics can be adjusted independently by changing the propagation constant in the first and second components of the seed solution, respectively. In Fig.~\ref{fig:long} we show metastable propagation of a relatively short necklace with $N=8$ which is composed of broader individual solitons confined within the flattop plateau of the same radius as above. The same figure   presents a   long necklace with $N=20$, hosted by a relatively broad flattop region.
 
 To conclude, we have presented a mechanism for metastable propagation of soliton necklaces forming in a two-component envelope and propagating in a media with competing cubic-quintic nonlinearity. The existence of metastable soliton necklaces explicitly exploits the combination of two key ingredients: first, the bimodal envelope provides effective intermodal attraction; second, the competing nonlinearity enables the existence of the flattop plateau, which suppresses the expansion of the necklace. Numerical simulations indicate metastable propagation over \rev{about a hundred} of diffraction lengths, even  when  weak random noise is added to the input envelopes. The number of solitons in metastable necklaces can be controlled by changing the width of individual solitons or the radius of the flattop plateau. These characteristics can be adjusted independently via the   propagation constants in the first and second components. In contrast to other examples of quasistationary necklaces in vectorial settings \cite{DesyatnikovOL01,Desyatnikov_JOSA02}, in our case the necklaces with the same number of solitons are present in both components. The adjacent solitons are out of phase in one component and in phase in the other. We expect that similar metastable solutions can also exist in other non-Kerr media, such as those with saturable nonlinearities.  A similar stabilization mechanism could potentially be exploited for clusters of quantum droplets \cite{Kartashov2019,Dong2026}  in binary atomic mixtures  with competing inter- and intra-species interactions. 
 
The stability results in this study are based on direct numerical propagation. Additional  confirmation of the existence of metastable necklaces could potentially be obtained via a linear stability analysis. However, such an analysis would be computationally costly, especially since the instability increments of metastable solutions are expected to be small and would therefore require high numerical accuracy to be computed reliably. We therefore  leave this issue for future studies. 
 
\textbf{Funding.}   Russian Science Foundation, Grant No. 26-11-00021.

\end{document}